# 48 channels 100-GHz tunable laser by integrating 16 DFB lasers with high wavelength-spacing uniformity

Zhirui Su[1], Rulei Xiao[1,*], Zhenxing Sun[1], Zijiang Yang[1], Yinchao Du[1], Zhao Chen[2], Jilin Zheng[1], Yunshan Zhang[1], Jun Lu[1], Yuechun Shi[1,4], Yi-Jen Chiu[3], and Xiangfei Chen[1]

[1]*Key Laboratory of Intelligent Optical Sensing and Manipulation of the Ministry of Education & National Laboratory of Solid State Microstructures & College of Engineering and Applied Sciences & Institute of Optical Communication Engineering, Nanjing University, Nanjing 210093, Jiangsu, China*
[2]*School of Electronic and Electrical Engineering, Wuhan Textile University, Wuhan, 430073, P.R. China*
[3]*Institute of Electro-Optical Engineering and Semiconductor Technology Research Development Center, National Sun Yat-Sen University, Kaohsiung 80424, Taiwan, R.O.C.*
[4]*Nanjing University (Suzhou) High-Tech Institute, Suzhou 215123, China*
**xrl@nju.edu.cn*

**Abstract:** We report a 48-channel 100-GHz tunable laser near 1550 nm by integrating 16 DFB lasers. High wavelength-spacing uniformity is guaranteed by the reconstruction-equivalent-chirp technique, which enables a temperature tuning range below 20°C.

## 1. Introduction

Wide-band tunable lasers have been employed in the optical communication and sensing areas, such as the reconfigurable DWDM system, distributed fiber sensor, and optical coherence tomography. According to L. A. Coldren's talk [1], there are currently two types of widely-used commercial tunable lasers. One is based on the Vernier-tuned distributed Bragg reflector structure and the other is based on the multi-wavelength laser array (MLA). For the Vernier-effect-based tunable lasers, wavelength control is an inherent issue that "mode hops" occurs, resulting in low mode stability and long calibration time [2]. For the MLA-based tunable lasers, the lasing mode is very stable owing to the DFB structure, but the DFB lasers in the MLA should have high yield and high wavelength-spacing uniformity. In addition, in the MLA-based tunable laser, an integrated/hybrid MMI or a funnel combiner is usually integrated for a single output waveguide [3]. This combiner will introduce a large intrinsic power loss, which increases with the laser count.

We propose a wide-band tunable laser by integrating 16 multi-wavelength DFB lasers. The wavelength-spacing uniformity is guaranteed by the reconstruction-equivalent-chirp (REC) technique, based on which 60-wavelength laser arrays have been demonstrated with high wavelength-spacing uniformity in our previous work [4]. The REC technique can equivalently realize the $\lambda/4$ phase shift and even arbitrary grating structure without using the electron beam lithography. Only two steps, one holography exposure and one mask exposure, are required when fabricating the REC-based sampled grating. The higher-order channel of the sampled grating is used to select the lasing wavelength. Compared to electron beam lithography, the REC technique can tune the sampling pattern on a much larger scale, thus resulting in a larger fabrication-error tolerance, which leads to precision integration of MLAs. The single-longitudinal-mode yield is guaranteed by the equivalent $\lambda/4$ phase shift in each DFB laser. 48 channels with 100-GHz-spacing near 1550 nm are realized that the temperature tuning range is lower than 20°C, from 19.3 to 37.3 °C.

## 2. Device structure and fabrication

The proposed laser mainly consists of four layers of different materials: (from up to down) a *p*-type InP contact layer, a grating layer, a separate confinement heterostructure multi-quantum well layer, and an *n*-type InP contact layer. These material layers are grown on an InP substrate by the metal-organic chemical vapor deposition technology. The equivalent phase-shifted sampled grating is defined by two steps of exposure, one holography exposure, and one sampled pattern mask exposure, and then etched by inductively coupled plasma etching. The ridge-type waveguides are formed by mask exposure and inductively coupled plasma etching. In the same waveguide, the electrical isolations between different lasers are enabled by shallow-etched grooves. Ti/Pt/Au metal layers are deposited as *p*- and *n*-electrodes by magnetron sputtering. The lift-off technique is used to define the electrode shape. Thermal annealing is then applied to improve metal contact and lower electrical resistance. Then the lasers are obtained after wafer cleaving, and two cleaved facets are coated with silicon oxide layer for anti-reflection and protection. Noted that heterogeneous integration techniques are not required, which makes the fabrication simple and low-cost.

The microscopic top view of the proposed tunable laser is shown in Fig. 1(a). The device consists of 16 DFB laser diodes (LDs), an active optical combiner, and a semiconductor optical amplifier (SOA). The 16 LDs are arranged as a 4-by-4 matrix. Each DFB laser has a $\lambda/4$ equivalent phase shift in the laser cavity. The wavelength spacings of the

LDs are designed as 2.4 nm. An "interleaving" scheme [3] has been used in the arrangement of the 16 Bragg wavelengths and the wavelength spacing of the neighboring two in-series lasers is 9.6 nm to reduce the crosstalk between the in-series lasers. The active optical combiner is realized by cascaded two-level Y-branches. The integrated SOA in front of the optical combiner is used for amplification and equalization of output power. The lengths of each LD, the combiner and the SOA are 450, 600 and 500 μm, respectively. The overall footprint of the chip is 3000×250 μm².

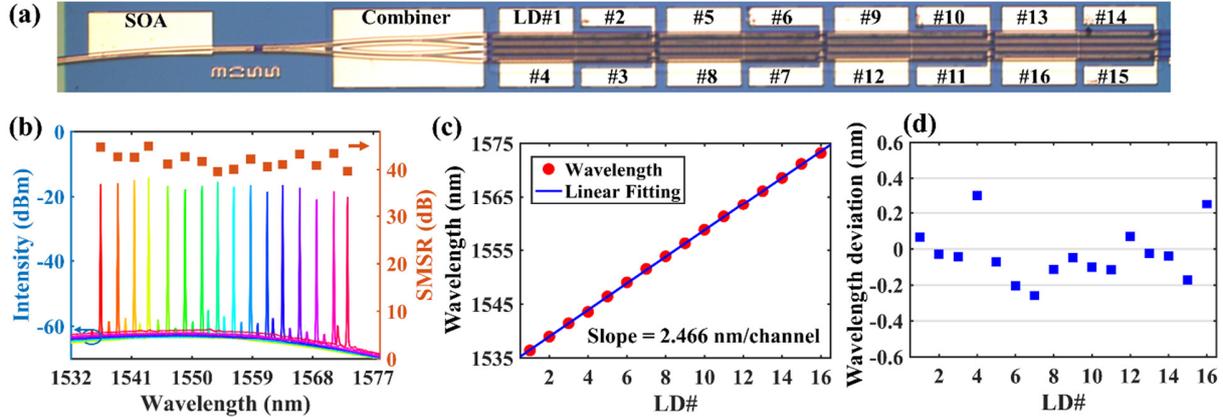

Fig. 1. (a) Microscopic top view of the proposed tunable laser. (b) Measured superimposed spectra of the 16 LDs (LD#1 to #16 from left to right). (c) Linear fitting and (d) deviations of the wavelengths of the 16 LDs. (LD: DFB laser diode)

### 3. Device characteristics

In the measurement, the laser chip was mounted on a sub-mount, the temperature of which was controlled at 20 °C via the TEC. The superimposed laser spectra of the 16 LDs are shown in Fig. 1(b). SMSRs of the 16 LDs are above 39.8 dB. The injection currents of the SOA ($I_{SOA}$), the combiner, the LDs ($I_{LD}$) and the in-front LDs (Transparent current, $I_{tr}$) are 40, 40, 70, and 20 mA, respectively. After the linear fitting of the wavelengths, a slope of 2.466 nm/channel is obtained as shown in Fig. 1(c). The wavelength deviations of 13 LDs are within ±0.2 nm and those of all the LDs are within ±0.4 nm.

We also measured the threshold current of each LD, as shown in Fig. 2(a). Threshold currents of all the LDs are approximately 30 mA. The threshold currents of LD#14, #15 and #16 are higher than others because the lasing wavelengths are far away from the peak of the material gain. Based on our statistics from 20 chips, the wavelength deviations of 89.3% LDs are within ±0.2 nm, and all the slopes are between 2.44 and 2.50 nm/channel, as shown in Fig. 2(b) and (c).

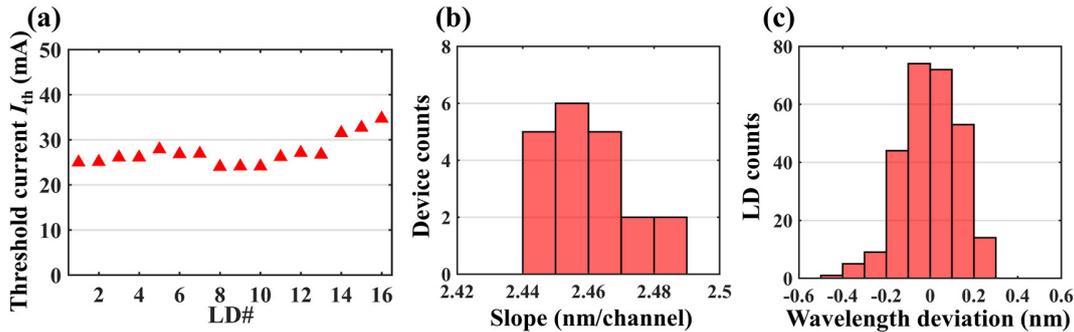

Fig. 2. (a) Measured threshold currents $I_{th}$ of the 16 LDs in a single tunable laser. Statistics of the wavelength slopes of 20 chips and wavelength deviations of all the LDs.

To find the optimal transparent current, we recorded the optical intensity and the optical SMSRs of all the four in-series LDs (LD#1, #5, #9, #13) under different transparent currents. As shown in Fig. 3(a), the optical intensity is balanced for four channels when the transparent current is approximately 22.0 mA. But the SMSRs are getting worse when the transparent current grows above 20.0 mA. As a result, the optimal transparent current is regarded as 20 mA and applied in the subsequent measurements. Then we measured the output power of four in-series lasers with respect to the $I_{SOA}$. All the four LDs can have an output power above 4.0 mW when the $I_{SOA}$ is larger than 70 mA.

By changing the temperature of the TEC, 48 channels of 100-GHz are obtained. The superimposed spectra are shown in Fig. 4(a). The SMSRs of all channels are larger than 39.8 dB. The temperature is tuned from 19.3 to 37.3 °C, with a range of only 18 °C. The temperature of each channel is shown in Fig. 4(b). By adjusting the current injected into the SOA, the output power of all the channels can be balanced. The superimposed spectra of power-balanced channels are shown in Fig. 4(c). Figure 4(d) shows the measured 20-dB linewidth of all the 48 channels, which are near 0.68 nm.

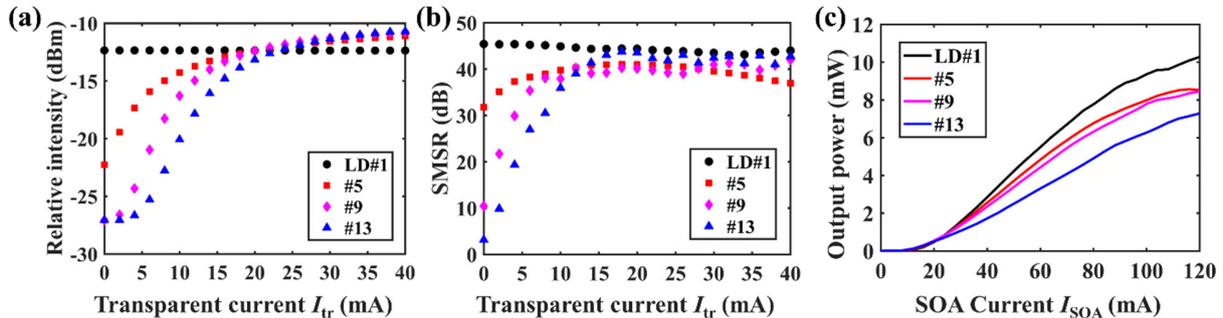

Fig. 3. Measured (a) relative optical intensities and (b) SMSRs of the in-series 4 LDs (LD#1, #5, #9, and #13) when the $I_{tr}$ is varied from 0 to 40 mA. (c) Measured output power of the 4 LDs when the $I_{tr}$ is 20 mA and the $I_{SOA}$ is varied from 0 to 120 mA.

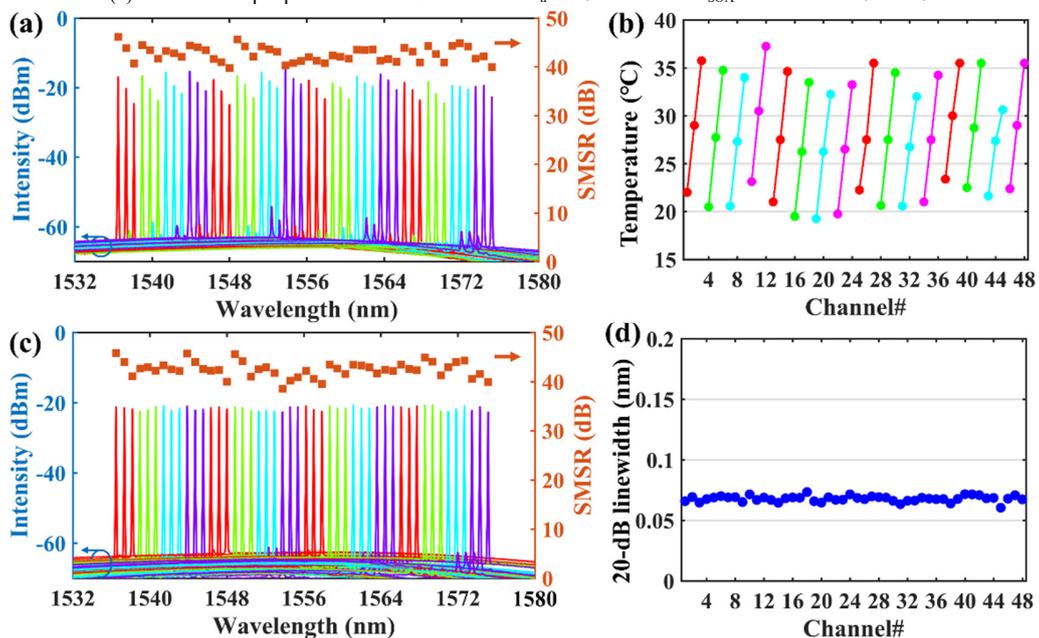

Fig. 4. Measured superimposed spectra and SMSRs of all 48 channels in the condition of (a) unbalanced and (c) balanced output power. (b) Tuning temperature and (d) measured 20-dB linewidth for all 48 channels.

## 4. Conclusion

We demonstrated a wide-band tunable laser by a high-density integration of 16 DFB laser diodes. The high wavelength-spacing uniformity is guaranteed by the REC technique. Based on our statistics from 20 tunable laser chips, wavelength deviations of 89.3% LDs are within ±0.2 nm. 48 channels with 100-GHz-spacing can be obtained by a temperature tuning range of 18 °C.

**Funding**

Chinese National Key Basic Research Special Fund (2017YFA0206401); Science and Technology Project and Natural Science Foundation of Jiangsu province (BE2017003-2, BK20160907); National Natural Science Foundation of China (61435014 and 11574141); Suzhou technological innovation of key industries (SYG201844).